\documentclass[aps,pra,twocolumn,superscriptaddress]{revtex4}

\usepackage{graphicx}
\usepackage{amssymb}
\usepackage{subfigure}
\usepackage{amsmath}
\usepackage{amsfonts}
\usepackage{bm}
\usepackage{color}

\newcommand{\ket}[1]{\ensuremath{|{#1}\rangle}}

\newcommand{\bra}[1]{\ensuremath{\langle{#1} |}}

\newcommand{\oper}[1]{\mathbf{\mathsf{#1}}}

\newcommand{\beq}{\begin{equation}}
\newcommand{\eeq}{  \end{equation}}
\newcommand{\bea}{\begin{eqnarray}}
\newcommand{\eea}{\end{eqnarray}}
\newcommand{\beqn}{\begin{eqnarray}}
\newcommand{\eeqn}{\end{eqnarray}}
\newcommand{\bit}{\begin{itemize}}
\newcommand{\eit}{  \end{itemize}}

\begin{document}

\title{Characterization of a spatial light modulator as a polarization quantum channel}

\author{G. Barreto Lemos}
\email[]{gabriela.barreto.lemos@univie.ac.at}
\affiliation{Instituto de F\'{\i}sica, Universidade Federal do Rio de Janeiro, Caixa Postal 68528, Rio de Janeiro, RJ 21941-972, Brazil}
\affiliation{Institute for Quantum Optics and Quantum Information, Austrian Academy of Sciences, Boltzmanngasse 3, Vienna A-1090, Austria}
\affiliation{Vienna Center for Quantum Science and Technology (VCQ), Faculty of Physics, University of Vienna, A-1090 Vienna, Austria.}
\author{J. O. de Almeida }
\author{S. P. Walborn}
\author{P. H. Souto Ribeiro}
\author{M. Hor-Meyll}
\affiliation{Instituto de F\'{\i}sica, Universidade Federal do Rio de
Janeiro, Caixa Postal 68528, Rio de Janeiro, RJ 21941-972, Brazil}

\begin{abstract}

Spatial light modulators are versatile devices employed in a vast range of applications to modify the transverse phase or amplitude profile of an incident light beam. Most experiments are designed to use a specific polarization which renders optimal sensitivity for phase or amplitude modulation. Here we take a different approach and apply the formalism of quantum information to characterize how a phase modulator affects a general polarization state. In this context, the spatial modulators can be exploited as a resource to couple the polarization and the transverse spatial degrees of freedom. Using a quasi-monochromatic single photon beam obtained from a pair of twin photons generated by spontaneous parametric down conversion, we performed quantum process tomography in order to obtain a general analytic model for a quantum channel that describes the action of the device on the polarization qubits.
We illustrate the application of these concepts by demonstrating the implementation of a controllable phase flip channel. This scheme can be applied in a straightforward manner to characterize the resulting polarization states of different types of phase or amplitude modulators and motivates the combined use of polarization and spatial degrees of freedom in innovative applications.

\end{abstract}


\maketitle
\section{Introduction}
Spatial light modulators (SLM) are ever more popular devices that have recently been employed for phase modulation of a light beam in a vast variety of classical and quantum optics~\cite{Moreno:12, meshulach98, Yao:06, Fatemi:07}, atom optics~\cite{McGloin:03}, optical tweezers~\cite{Curtis2002169, grier2003}, quantum chaos~\cite{lemos2012}, quantum metrology~\cite{PhysRevLett.88.203601}, quantum information~\cite{PhysRevLett.98.083602, PhysRevA.86.052327, Abouraddy-slm-2012, Fickler02112012, lima-slm-teleportation2013, Leach06082010, MUBS}, and quantum communication experiments~\cite{Gibson:04, Gruneisen:08}. The increasing popularity  of this device is mainly due to its versatility: SLMs have been used to produce different orbital angular momentum (OAM) states of light; they can act as digital lenses or holograms,  tunable filters, among other applications.
\par
The phase and amplitude modulation depends strongly on the wavelength of the incident light and it is well known that reflective SLM's have optimal functionality for a specific linear polarization~\cite{moreno2003}. Once it has been calibrated for phase and/or amplitude modulation of a monochromatic light beam, the SLM constitutes an automated high resolution mask that can easily be programed and has a fast response. 
\par
Recently, experiments were reported in which polarization and transverse spatial degrees of freedom (DOF) of photons are simultaneously addressed in a variety of applications~\cite{Marruci2006,Nagali2010}. An SLM was used in Ref.~\cite{Moreno:12}, to generate several interesting polarization states by modulating the transverse phase distribution of light. 
\par 
Up to now few experiments have exploited the SLM as a means to create entanglement between polarization and spatial DOF of photons. In Ref.~\cite{Abouraddy-slm-2012} the polarization of light was used as a control qubit in a three-qubit quantum gate, where the other two qubits are the spatial parity in Cartesian directions. In Ref.~\cite{lemos2012} the effects of a  chaotic wavefront on a polarization qubit was investigated, while in Ref.~\cite{Lemos:14} an optical integration method was demonstrated. In a very recent article~\cite{hormeyll14}, entanglement between transverse spatial variables of a pair of twin photons was detected through polarization measures. These diverse examples provide enough evidence of the usefulness of the coupling between the polarization and the transverse
spatial DOF of a light beam that is available through the use of SLMs.
In order to obtain advantage of the vast range of possibilities provided by the SLM to control the polarization and spatial properties of light, it is necessary to characterize this process.
\par 
In this article we present a formal description of the action of an SLM on a general polarization state of an incident light beam using the quantum process framework~\cite{NielsenChuang}. In order to make a direct connection to quantum information experiments and to emphasize how this device could be used in these applications, we
characterize the action of the SLM as a noisy quantum channel acting on a qubit encoded in the polarization degree of freedom of a photon. We employ quantum process tomography (QPT) to obtain the Kraus operators associated with the noisy quantum channel for different patterns displayed on the SLM screen. We then show how the SLM can be used to implement a phase flip quantum channel~\cite{NielsenChuang} 
in a straightforward and controllable manner.  Thus the SLM can be easily incorporated into photonic experiments involving the study of quantum properties such as entanglement and decoherence~\cite{almeida07, farias09, farias12}, especially when one would like to couple spatial and polarization degrees of freedom~\cite{Fickler13}. 
\par
\section{Experimental Setup}

\begin{figure}
\begin{center}
 \includegraphics[width=8cm]{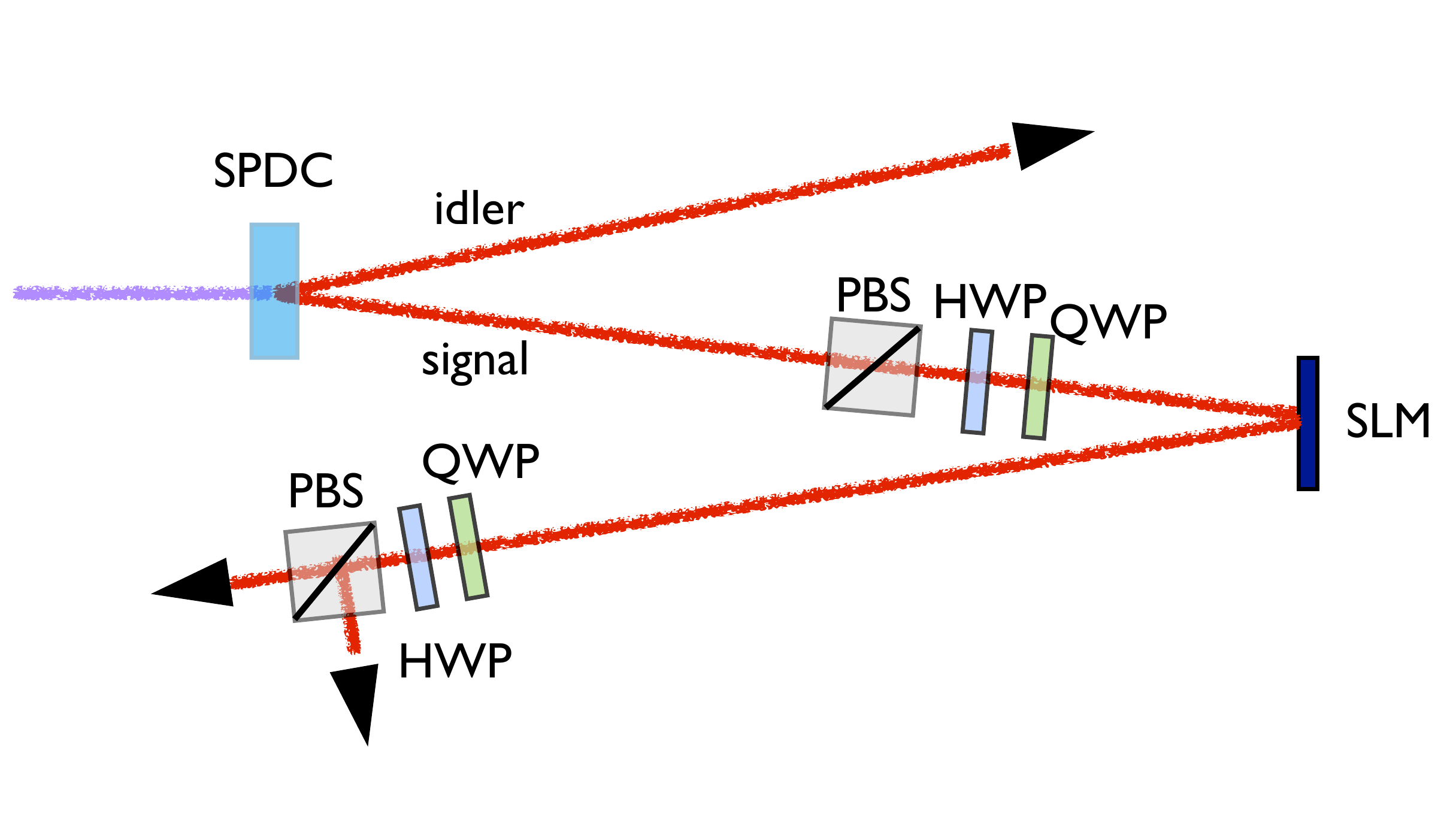}
    \end{center}
  \caption{Experimental setup: a polarizing beam splitter (PBS), a half-wave plate (HWP), and a quarter-wave plate (QWP) are used to prepare four polarization states of light, $|H\rangle$ (horizontal polarization), $|V\rangle$ (vertical polarization), $|D\rangle\equiv(|H\rangle+|V\rangle)/\sqrt{2}$, and $|R\rangle\equiv(|H\rangle+i|V\rangle)/\sqrt{2}$. The polarized photon reflects after nearly normal incidence upon the SLM screen, which displays an image with a certain gray level $g$ ranging from $0$ to $255$. A QWP, a HWP and a PBS are used to perform projective measurements in $|H\rangle, |V\rangle, |D\rangle, |R\rangle$ states using a 
 single photon detector.
}\label{fig:setup}
\end{figure}

The experimental setup is shown in Fig.~\ref{fig:setup}.
We use a $325$
nm He-Cd laser incident on a Beta Barium Borate (BBO) crystal to produce twin degenerate photons at $650$
nm via type-I spontaneous parametric down-conversion
(SPDC). The idler photon is sent directly to a detector and is used only for heralding the signal photon. 
A polarizing beam splitter (PBS), a half-wave plate (HWP), and a quarter-wave plate (QWP) are used to prepare the initial state of the signal photon, incident upon a Pluto reflective phase-modulation SLM manufactured by Holoeye Photonics, with resolution $1920 \times 1080$ pixels and $8\;\mu$m pixel pitch. This SLM is essentially composed of a programmable LCD screen in which one can display any $8$-bit ($256$ gray levels) image. 
After reflection on the SLM, the signal photon is sent to a polarization detection system. Lenses (not shown) image the SLM plane onto the detection plane. A QWP, a HWP, and a PBS are used to realize projective measurements in different polarization states, and coincidence photon counting is performed between signal and idler single-photon detectors.

\section{The action of the SLM as a quantum process acting upon the polarization qubit}

\begin{figure}\begin{center}   
 \includegraphics[width=8cm]{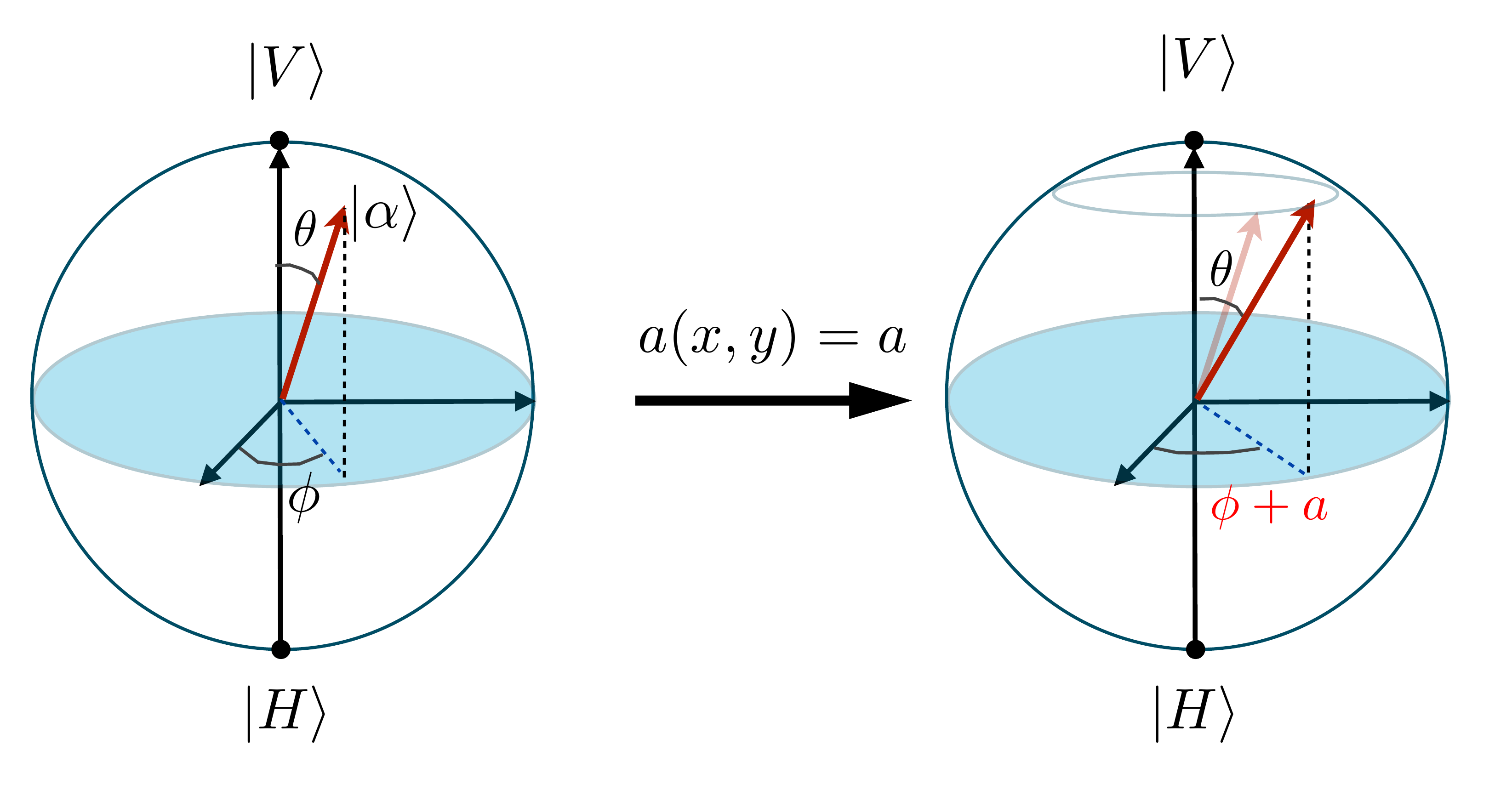}
    \end{center}
  \caption{Bloch sphere representation of the ideal action of the SLM on a qubit encoded in the polarization state of the incident light. For a constant function $g(x,y) =g$, the SLM would ideally imprint a constant phase $a(x,y)=a=2\pi g/255$ upon the horizontal polarization component, and the vertical polarization component of light would remain unchanged. This is represented by a rotation of the polarization qubit around the vertical axis in the Bloch sphere by the angle $a$.}
  \label{fig:ideal-bloch}
\end{figure}

\par
We use a SLM previously calibrated for linear phase-only modulation on horizontally polarized light at $650$nm wavelength. 
Expressed in terms of a quantum evolution in the space defined by polarization and transverse spatial DOF, the ideal action of the SLM can be described by the operator 
\begin{equation}
 \oper{S} = \ket{H} \bra{H}  \otimes \oper{U} + \ket{V}\bra{V} \otimes \oper{I},
 \label{S_oper}
 \end{equation}
where $|H\rangle$ ($|V\rangle$) represents the horizontal (vertical) polarization state, operator $\oper{I}$ is the identity, and $\oper{U}$ implements the spatial phase modulation such that
\begin{eqnarray}
\oper{U} = e^{i a(\oper{x},\oper{y})}\equiv e^{i(2\pi/255) g(\oper{x},\oper{y})},
\end{eqnarray}
 where $g(x,y)$ is an arbitrary function ranging from $0$ to $255$ which specifies the degree of modulation
for the pixel at position $(x,y)$ on the SLM screen.  Representing the transverse spatial state of light by $|\psi\rangle$, application of the operator $\oper{S}$ can be represented by the following quantum map:
\begin{eqnarray}\label{eq:ideal-map}
|\psi\rangle |H\rangle& \rightarrow & e^{ia(\oper{x},\oper{y})}|\psi\rangle |H\rangle,\nonumber\\
|\psi\rangle |V\rangle & \rightarrow & |\psi\rangle |V\rangle.
\end{eqnarray}
For a constant function $g(x,y) =g$, the SLM would ideally imprint a constant phase $a(x,y)=a=2\pi g/255$ on the horizontal polarization component, while the vertical polarization component would remain unchanged. In this case, the action of the SLM on the state $|\psi\rangle|\alpha\rangle$ where $|\alpha\rangle$ is an arbitrary pure polarization state 
\begin{eqnarray}
|\alpha\rangle =\cos(\theta/2)|V\rangle +e^{i\phi}\sin(\theta/2)|H\rangle, 
\label{alpha}
\end{eqnarray}
is given by
\begin{eqnarray}
\oper{S}|\psi\rangle|\alpha\rangle=|\psi\rangle[\cos(\theta/2)|V\rangle +e^{i(\phi+a)}\sin(\theta/2)|H\rangle].
\end{eqnarray}
 In the Bloch sphere  representation of the polarization qubit, this would mean a rotation of $|\alpha\rangle$ by an angle $a$ around the vertical axis, as shown in Fig.~\ref{fig:ideal-bloch}. It is worthwhile to note that a constant phase does not entangle the polarization and the transverse spatial degree of freedom. We will see in Sec.~\ref{general} that a nonuniform phase profile can create entanglement between these degrees of freedom.
 
 \par
\begin{figure}
\centering   
\includegraphics[scale=0.35]{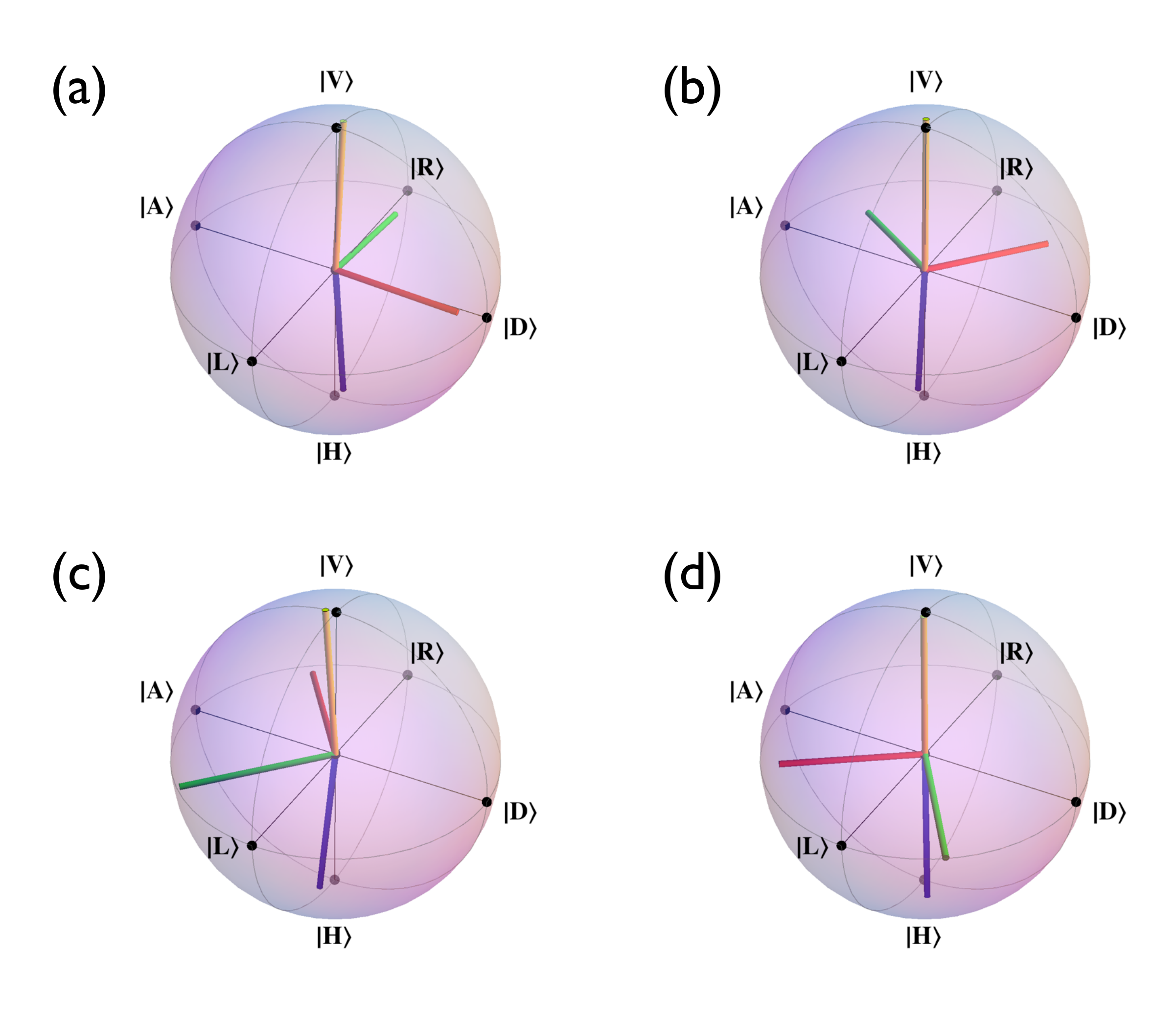}
\caption{Bloch sphere representation of the output states after a polarization qubit is prepared in four different input states and then reflected by the SLM programed with a constant phase profile. The vectors correspond to output states when the initial states were $\ket{H}$ (blue), $\ket{V}$ (orange), $\ket{D}$ (red), and $\ket{R}$ (green). 
The different spheres correspond to constant phase modulations associated with the following gray values: (a) $g=0$,  (b) $g=255/8$, (c) $g=3\times 255/8$, and (d) $ g=5\times 255/8$.}
\label{fig:spheres}
\end{figure}

In our experiment, we prepared the polarization states $|H\rangle, |V\rangle, |D\rangle\equiv (|H\rangle+|V\rangle)/\sqrt{2}, |R\rangle\equiv (|H\rangle+i|V\rangle)/\sqrt{2}$, and in each case we measured the output polarization state after reflection upon the SLM.  
We analyze the action of the SLM on each of the above states when a constant image is displaced on the LCD screen. This is illustrated in the Bloch sphere, as shown in Fig.~\ref{fig:spheres}, for (a) $g=0$, (b) $g=255/8$, (c) $g=3\times 255/8$, and (d) $g=5\times 255/8$. Each colored vector corresponds to one output state after modulation
of a given input state, according to the correspondence:
blue vector $=$ input state $|H\rangle$, orange vector $=$ 
input state $|V\rangle$, red vector $=$ input state $|D\rangle$, and 
green vector $=$ input state $|R\rangle$.

One can observe from  Fig.~\ref{fig:spheres} that when the SLM is on, the programed rotation around the vertical axis is accompanied by an unexpected shrinking of the length of the Bloch vector, which is due to a reduction of the purity (loss of coherence) of the polarization state. The purity of a state $\rho_f$  is defined as ${\rm Tr} \rho_f^2$, which can range from $0.5$ (maximally mixed state) to $1$ (pure state). The purity of the final polarization state associated to $\ket{H}$ and  $\ket{V}$ input states (blue and orange vectors in  Fig.~\ref{fig:spheres}) is $ \approx1$, while for $\ket{D}$ and  $\ket{R}$ input states the final purity is in the range $[0.83,0.96]$.

We can interpret the overall evolution of the polarization qubit due to reflection on the SLM as a quantum process -- the transverse spatial DOF acting as an environment -- which can be written in the operator sum representation
\begin{equation}
\mathcal{M}(|\alpha\rangle\langle \alpha |)=\oper \sum_{i=0}^{3} M_i|\alpha\rangle\langle \alpha| M_i ^\dagger ,
\end{equation} 
where  $M_i$  are the Kraus operators that describe the process including the noisy effect mentioned above, and normalization of the resulting state implies that 
\begin{equation}
\sum_{i=0}^{3} M_i^\dagger M_i =1.
\end{equation} 
 Therefore, in order to obtain complete information about the effect of the SLM upon the qubit encoded in the polarization, we performed a standard quantum process tomography~\cite{NielsenChuang} for images corresponding to a uniform phase $a=n\pi/4$, with the integer $n$ varying from $0$ to $7$. 

 Our process tomography shows that the action of the SLM on the polarization degree of freedom can be suitably represented by the following 
 Kraus operators, which correspond to a phase flip channel~\cite{NielsenChuang}, characterized by the parameter $p$, combined with the programed $a$-phase rotation:  
\begin{eqnarray}
M_0 &=&\sqrt{1 - p} \left(\begin{array}{cc}e^{i a}& 0\\ 0&1 \end{array}\right),\nonumber\\
M_1&=&\sqrt{p} \left(\begin{array}{cc}e^{i a}& 0\\0& -1\end{array}\right),\nonumber\\
M_2&=&M_3=\left(\begin{array}{cc}0& 0\\0& 0\end{array}\right),
\label{eq:nonideal-map}
\end{eqnarray}
in the basis $|H\rangle=(1\;0)^T$ and $|V\rangle=(0\;1)^T$.
To obtain the value of $p$ that best fits the action of our SLM on the polarization degree of freedom and the corresponding average fidelity $F$ of our experimental Kraus operators compared to the ideal ones described in Eq.~(\ref{eq:nonideal-map}) we proceed as follows.

As stated by Jamiolkowski  isomorphism~\cite{Alberwerner} 
there is a duality between quantum channels and quantum states, described by density matrices. This isomorphism can be used to obtain the fidelity between the experimental and theoretical quantum channels of Eq.~(\ref{eq:nonideal-map}) by means of the fidelity between their corresponding dual quantum states.  We follow this prescription and, for each SLM image corresponding to a uniform phase, 
we obtained the dual states for the experimental and theoretical quantum channels applying the corresponding Kraus operators in the second qubit of an initially maximally entangled two-qubit state. 
For each uniform phase we numerically calculated the value of $p$ that maximizes the fidelity $F={\rm Tr} {\sqrt{\sqrt{\rho}\;\sigma(p)\;\sqrt{\rho}}}$ between the experimental and theoretical dual density matrices, $\rho$ and $\sigma(p)$, respectively. We then calculated the average value of $p$ and $F$ as well as their corresponding standard deviations, obtaining $\overline{p}=0.08\pm 0.02$ and $\overline{F} =0.998\pm0.002$. 

In order to verify that the shrinking (loss of coherence) was caused by the SLM, we realized QPT without the SLM.  This should be equivalent to doing QPT of an identity channel.  We observed a slight shrinking of the Bloch vectors with purities ranging from 0.95 to 0.99, which we attribute to imperfections in the wave plates.  The slight decrease in purity is completely consistent with polarization interference curves obtained with the half- and quarter-wave plates used in the tomography process, which gave visibilities of $0.982\pm0.005$ and $0.316\pm0.005$, where ideal values should be $1$ and $1/3$, respectively.  To investigate the effect of these slight imperfections on the QPT , we evaluated how similar the experimental identity channel was to a phase flip channel. Using the procedure described above to maximize the fidelity between the phase flip channel and the experimental identity, but now including the phase $a$ of Eq.~(\ref{eq:nonideal-map}) as an adjustable parameter, we achieved a fidelity $F=0.997$ for $p=0.026$ and $a=0.07$. Though this demonstrates that our QPT system includes an effect similar to the one produced by a phase flip channel, the value of $p$ obtained is quite distinct from what we achieve with the SLM in the setup.  In that case, we obtain near unity fidelity with $p=0.08$ on average, indicating that the SLM is primarily responsible for the loss of coherence produced by the phase flip channel described by Eq.~(\ref{eq:nonideal-map}). 

\begin{figure}
\centering   
\includegraphics[scale=0.22]{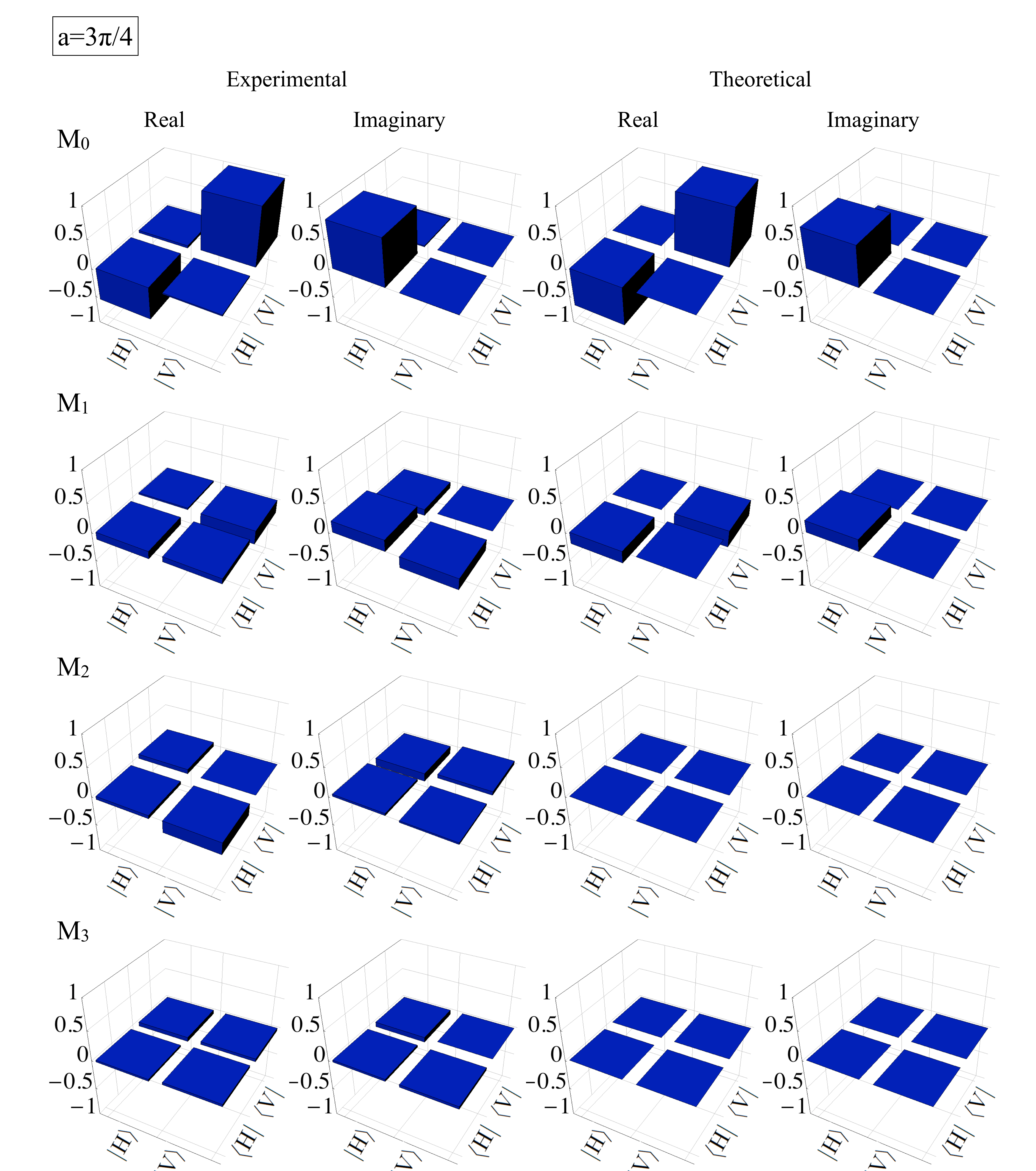}
\caption{Experimental and theoretically predicted Kraus operators for a user-defined constant phase $a=3 \pi/4$.}
\label{fig:Kraus1}
\end{figure}

\begin{figure}
\centering   
\includegraphics[scale=0.22]{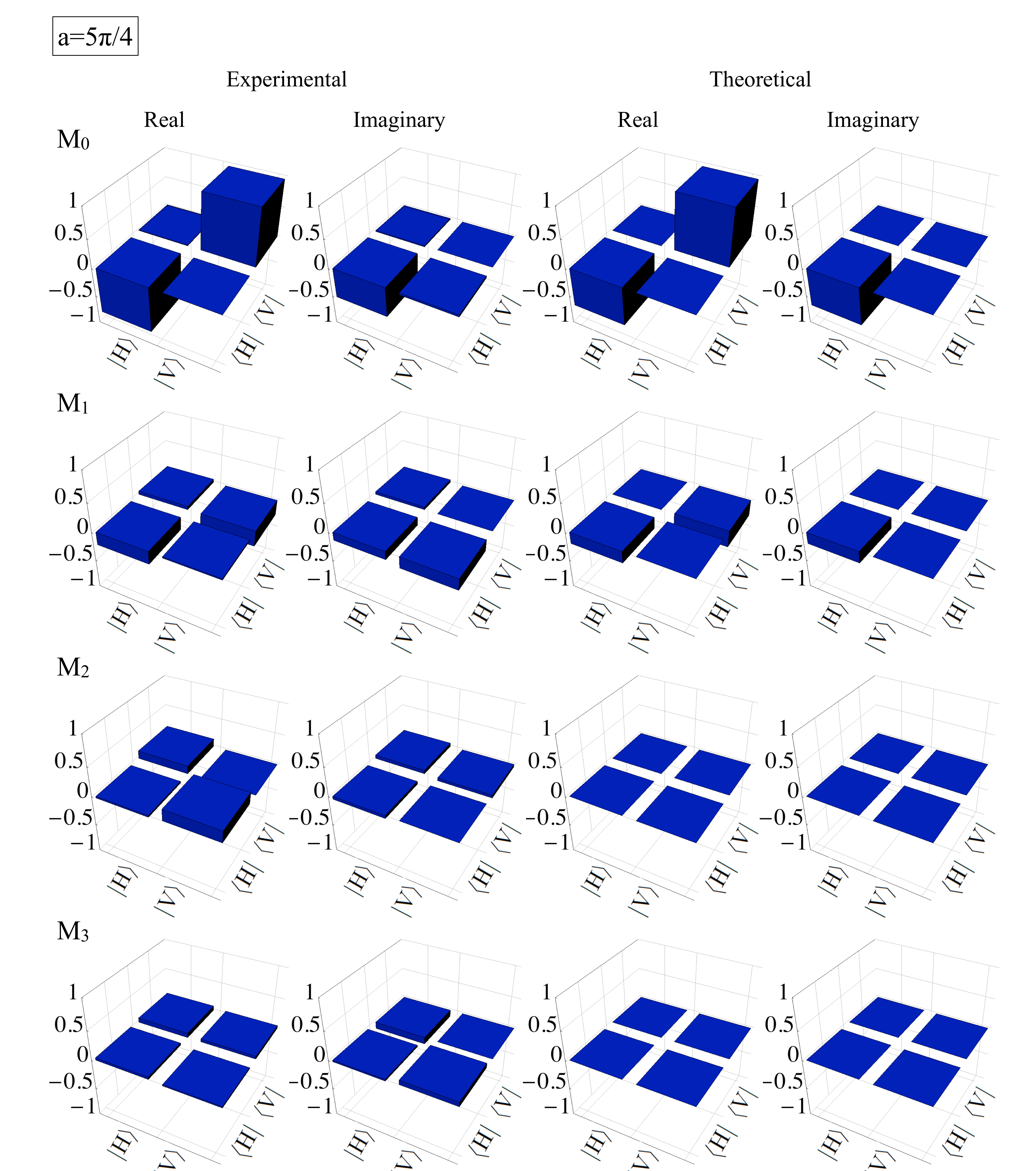}
\caption{Experimental and theoretically predicted Kraus operators for a user-defined constant phase $a=5 \pi/4$.}
\label{fig:Kraus2}
\end{figure}

In order to illustrate the agreement between the Kraus operators obtained experimentally through the tomography and the above proposed model [Eq.~(\ref{eq:nonideal-map}) with $p=0.08$], we show in  Figs.~\ref{fig:Kraus1} and \ref{fig:Kraus2}  the real and imaginary components of the corresponding matrices, for $a=3 \pi/4$ and $a=5 \pi/4$.
The Kraus operators shown in  Figs.~\ref{fig:Kraus1} and \ref{fig:Kraus2} are associated with the vector transformations shown in Figs.~\ref{fig:spheres}(c) and \ref{fig:spheres}(d), respectively. By applying the Kraus operators in the initial pure state $\rho_{\alpha}=|\alpha\rangle\langle\alpha|$, we see that the parameter $p$ gives the loss of coherence in the final state $\rho_f$:
\begin{eqnarray}
\rho_f&=&\mathcal{M}(\rho_{\alpha})=\sum_{i=0}^{3}  M_i \left(\begin{array}{cc}\sin^2(\frac{\theta}{2})&e^{i \phi}\frac{\sin\theta}{2}\\ 
e^{-i \phi}\frac{\sin\theta}{2}&\cos^2(\frac{\theta}{2}) \end{array}\right)M_i^\dagger\nonumber\\
&=&\left(\begin{array}{cc}\sin^2(\frac{\theta}{2})& e^{i (\phi+a)}(1-2p)\frac{\sin\theta}{2}\\ 
e^{-i (\phi+a)}(1-2p)\frac{\sin\theta}{2}&\cos^2(\frac{\theta}{2}) \end{array}\right).\nonumber\\
\label{dephasing}
\end{eqnarray}
When an SLM device is used in quantum information applications, the relevance of the spurious decoherence effect that accompanies the $a$-phase rotation is determined by the value of $p$, which in general relies on particular SLM characteristics and can vary for different models. A complete decoherence process takes place for $p=0.5$, when off-diagonal terms in $\rho_f$ vanish.


\par
\section{Kraus operators in the general case}
\label{general}
Note that, in general,  the mask displayed on the SLM screen need not be uniform. Assuming  ideal action of the SLM, 
application of operator $\oper{S}$ of Eq.~(\ref{S_oper}) on an arbitrary state $|\psi\rangle|\alpha\rangle$, results in 
\begin{eqnarray}
\oper{S}(|\psi\rangle|\alpha\rangle)=\cos(\theta/2)|\psi\rangle|V\rangle +e^{i a(\oper{x},\oper{y})}e^{i\phi}\sin(\theta/2)|\psi\rangle |H\rangle,\nonumber\\
\label{axy}
\end{eqnarray}
where now $a(x,y)$ is an arbitrary function of the transverse spatial coordinates $(x,y)$.
In this case, it is possible to generate entanglement between the polarization and the spatial DOF.
Let us consider that $|\psi \rangle$ is prepared in the pure initial state given by 
\begin{equation}
\ket{\psi}= \iint \psi(x,y) |x\rangle |y\rangle \mathrm{d}x \mathrm{d}y,
\label{purepsi}
\end{equation}    
where $|x\rangle |y\rangle$  refers to a single photon in the position representation, with $x$ and $y$ corresponding to horizontal and vertical Cartesian coordinates~\cite{walborn10}.   
In order to obtain the Kraus operators in the general case we trace out the transverse spatial degree of freedom in Eq.~(\ref{axy}):
\begin{eqnarray}
\rho_p&=&\iint\!\langle x'|\langle y'| \oper{S}(|\psi\rangle|\alpha\rangle \langle\psi|\langle\alpha|)\ \oper{S}^\dagger|x'\rangle|y'\rangle \;  \mathrm{d}x'\; \mathrm{d}y'=\nonumber\\
&&\cos^2\left(\frac{\theta}{2}\right)|V\rangle\langle V|+\sin^2\left(\frac{\theta}{2}\right)|H\rangle\langle H|+\nonumber\\
&&e^{-i\phi} \frac{\sin \theta}{2}|V\rangle\langle H|\int e^{-i a(x,y)} |\psi(x,y)|^2  \;\mathrm{d}x\; \mathrm{d}y+\nonumber\\
&&e^{i\phi}\frac{\sin \theta}{2}|H\rangle\langle V|\int e^{i a(x,y)} |\psi(x,y)|^2  \;\mathrm{d}x\; \mathrm{d}y.
\label{traceout}
\end{eqnarray}

Representing Eq.~(\ref{traceout}) in  matrix notation, we have 
\begin{eqnarray}
\rho_p=\left(\begin{array}{cc}\sin^2(\frac{\theta}{2})&e^{i \phi} \langle e^{ia(x,y)}\rangle_{\psi}\frac{\sin\theta}{2}\\
e^{-i \phi} \langle e^{-ia(x,y)}\rangle_{\psi}\frac{\sin\theta}{2}&\cos^2(\frac{\theta}{2}) \end{array}\right),
\label{matrixaxy}
\end{eqnarray}
 where $ \langle e^{ia(x,y)}\rangle_{\psi}\equiv\int e^{i a(x,y)} |\psi(x,y)|^2  \;\mathrm{d}x \mathrm{d}y $ is the mean value of the imprinted phase in the transverse spatial state  $|\psi\rangle$ of the incident beam.

As an entanglement measure between polarization and spatial DOF of the state in Eq.~(\ref{axy}), we use the concurrence~\cite{wooter1998,PhysRevA.64.042315}:
\begin{eqnarray}
C = \sqrt{2(1-{\rm{Tr}} \rho_p^2)} = \sin\theta \sqrt{[1-|\langle e^{ia(x,y)}\rangle_{\psi}|^2]}.
\label{entropy}
\end{eqnarray}
By suitable choice of the $(x,y)$-dependent phase, one could obtain $| \langle e^{ia(x,y)}\rangle _\psi|<1$, creating entanglement   ($C>0$) between polarization and spatial DOF for any initial polarization state given by a superposition of $|H\rangle$ and $|V\rangle$, i.e., $\theta \neq0,\pi$ in Eq.~(\ref{alpha}). The maximum degree of entanglement is achieved for $\langle e^{ia(x,y)}\rangle_{\psi}=0$ and 
 $\theta = \pi/2$. 

Expressing the complex number $\langle e^{ia(x,y)}\rangle _\psi\equiv A_{\psi}e^{ia_{\psi}}$, and considering the phase flip channel
effect observed in the process  tomography, the polarization state $\rho_p$ takes the final form
\begin{eqnarray}
\rho_{pf}=\left(\begin{array}{cc}\sin^2(\frac{\theta}{2})&e^{i\phi_{\rm{eff}}}(1-2p_{\rm{eff}})\frac{\sin\theta}{2}\\ 
e^{-i\phi_{\rm{eff}}}(1-2p_{\rm{eff}})\frac{\sin\theta}{2}&\cos^2(\frac{\theta}{2}) \end{array}\right),\nonumber\\
\end{eqnarray}
where $(1-2p_{\rm{eff}})\equiv(1-2p)A_{\psi}$ and $\phi_{\rm{eff}}=\phi+a_{\psi}$.
Accordingly, the Kraus operators $M_0$ and $M_1$ of Eq.~(\ref{eq:nonideal-map}) generalize to
\begin{eqnarray}
M_0 &=&\sqrt{1 - p_{\rm{eff}}} \left(\begin{array}{cc}e^{ia_{\psi}}& 0\\ 0&1 \end{array}\right),\nonumber\\
M_1&=&\sqrt{p_{\rm{eff}}} \left(\begin{array}{cc}e^{ia_{\psi}}& 0\\0& -1\end{array}\right),
\label{eq:nonideal-map_general}
\end{eqnarray}
where 
\begin{eqnarray}
p_{\rm{eff}}=\frac{1-(1-2p)A_{\psi}}{2},
\label{peff}
\end{eqnarray}
while $M_2$ and $M_3$ are still given by Eq.~(\ref{eq:nonideal-map}).

\section{Implementing a controllable phase flip channel}

\begin{figure}
\centering    
\includegraphics[scale=0.3]{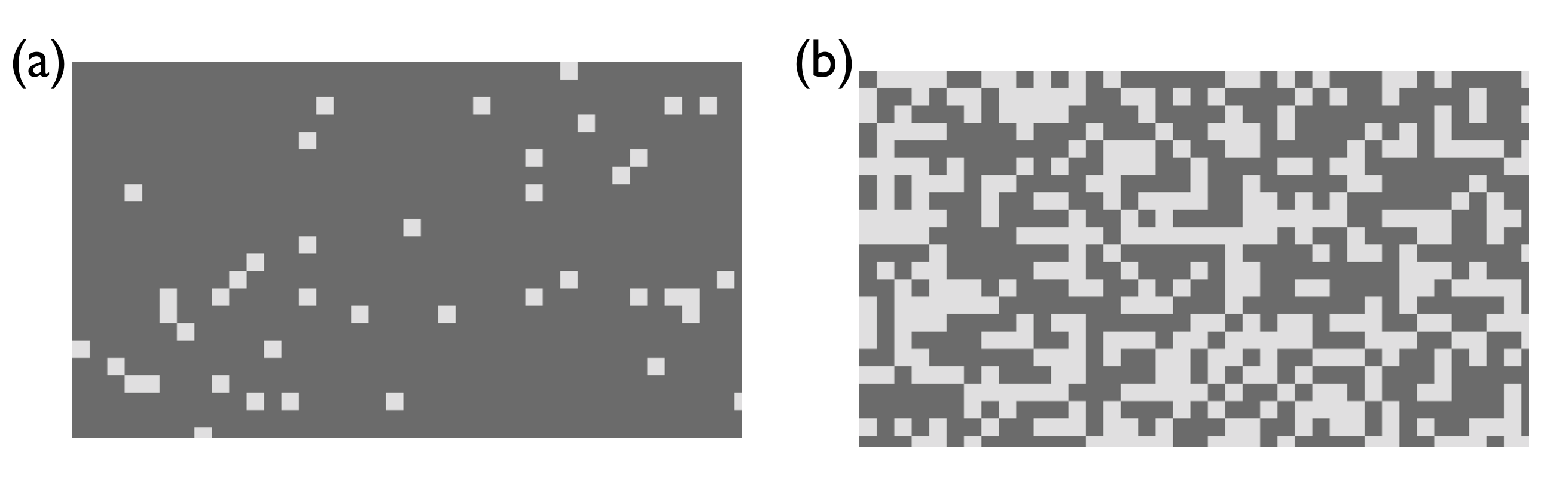}
 \caption{Screen shot of the phase mask produced on the SLM for the realization of a controllable phase flip channel for (a) $q=0.05$ and (b) $q=0.45$.}  
 \label{fig:phase-flip-mask}
\end{figure}

\par
After characterizing the intrinsic action of the SLM in the quantum channels formalism, we propose to use these results in quantum information experiments to implement a given quantum channel. As an example, we show in this section  how this device can be used to implement a controllable phase flip channel, given by the following Kraus operators: 

\begin{eqnarray}
E_0 &=&\sqrt{1-q} \left(\begin{array}{cc}1& 0\\ 0& 1\end{array}\right),\nonumber\\
E_1&=&\sqrt{q}\left(\begin{array}{cc}1& 0\\0& -1\end{array}\right),
\label{Kraus:phase-flip}
\end{eqnarray}
where the controllable parameter $q$ sets the degree of decoherence.

\par
For the sake of clarity, we first consider the implementation of the controllable phase flip channel assuming $p=0$. This is justified by the fact that although this parameter varies for different models and manufacturers, one in general expects it to be small and, as we will discuss in a moment, $p$ only plays a detrimental role when $q< p$.

To obtain the Kraus operators corresponding to a given value of q, we use a random number generator to produce a phase mask such that each square cell of $100 \times 100 $ pixels is filled with either $g=0$ (corresponding to zero-phase shift) or $g=255/2$ (corresponding to a $\pi$-phase shift), with probability $1-q$ and $q$, respectively. Figure~\ref{fig:phase-flip-mask} shows pictures of the SLM screen for (a) $q=0.05$ and (b) $q=0.45$. On average the mask implements a $\pi$-phase shift on $q\times100\%$ of the transverse spatial part of the horizontally polarized component of the incident beam. The vertically polarized component of the field receives no phase shift. The distribution of the modulated cells does not need necessarily to be
random. If we knew exactly the SLM area upon which the photon wavefront is incident, we could deterministically distribute the $\pi$ phase in $q\times100\%$ of the incidence area and the zero phase elsewhere. However, distributing the phase cells randomly eliminates prior need for this spatial information.

Assuming a near constant probability amplitude for the transverse spatial distribution $\psi(x,y)$ in Eq.~(\ref{purepsi}), for a given $q$-phase profile as described above we have
\begin{eqnarray}
\langle e^{ia(x,y})\rangle_\psi=1-2q\nonumber\\
\Rightarrow A_{\psi}=(1-2q), \; a_{\psi}=0.
\label{cpf}
\end{eqnarray} 
Plugging Eq.~(\ref{cpf}) in the Kraus operators described in  Eq.~(\ref{eq:nonideal-map_general}),  we find $p_{\rm{eff}}=q$ and $e^{ia_{\psi}}=1$, matching the Kraus operators  for the phase flip in Eq.~(\ref{Kraus:phase-flip}).

\begin{figure}
\centering    
 \includegraphics[scale=0.3]{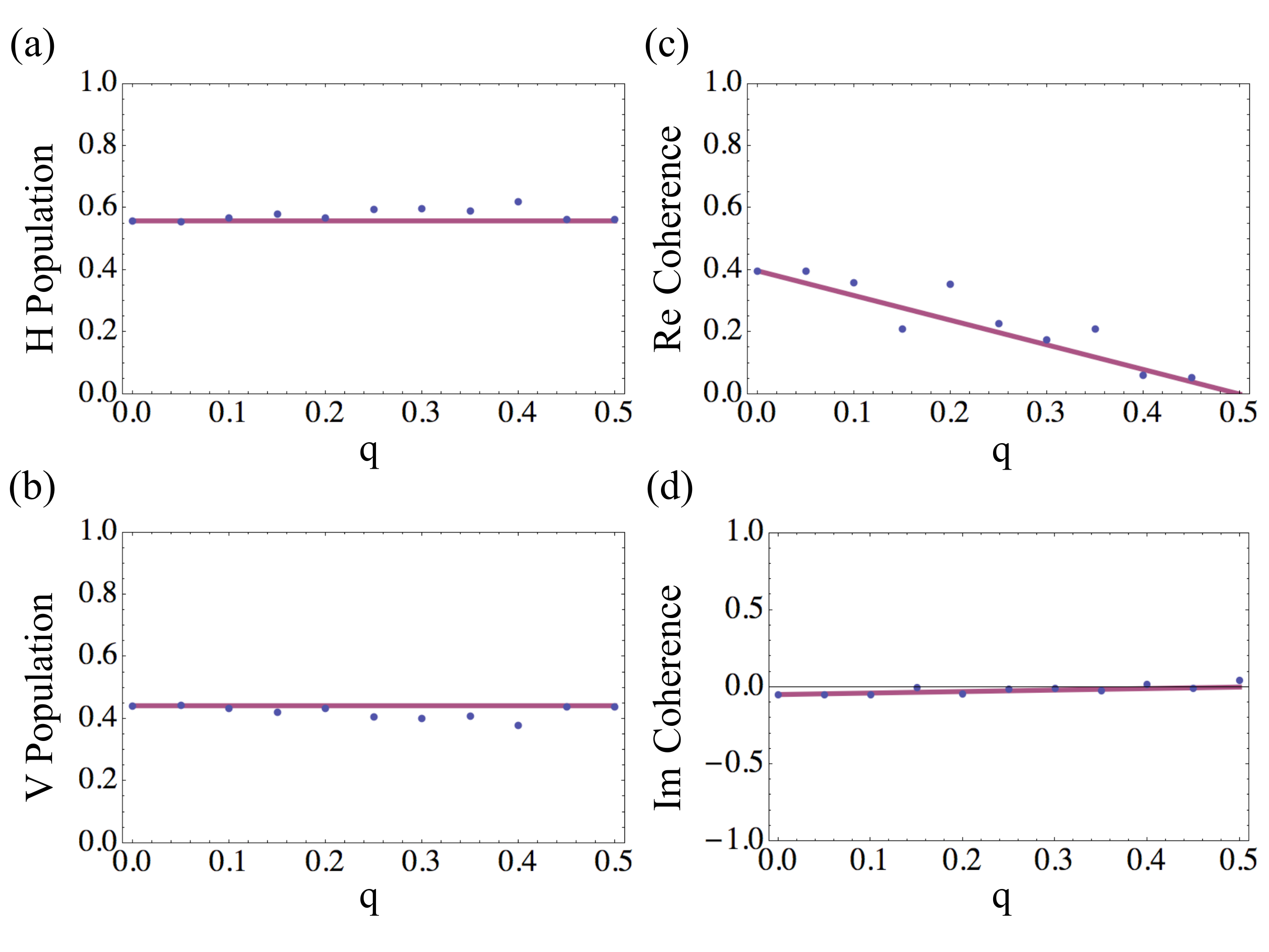}
   \caption{Final state of the polarization qubit, initially prepared in state $\ket{D}$, as a function of the parameter $q$ of the phase flip quantum channel. The blue dots show the experimental data for (a) population $H$, (b) population $V$, (c) real part of the coherence and, (d) imaginary part of the coherence. The red solid lines are theoretical curves of the corresponding quantities, obtained by evolving the initial experimental state with an ideal quantum map. Errors calculated by the Monte Carlo method led to errors bars smaller than the graph points.}\label{fig:phase-flip}
\end{figure}

To illustrate the action of the controllable phase flip channel, we prepared the $|D\rangle $ polarization state and performed  polarization measurements after reflection on the SLM. The experimental data (blue dots) plotted in Fig.~\ref{fig:phase-flip} show the final polarization state for different values of the channel parameter $q$. Figure~\ref{fig:phase-flip}(a)  shows the population $H$,
 Fig.~\ref{fig:phase-flip}(b)  shows the population $V$,  Fig.~\ref{fig:phase-flip}(c)  shows the real part of the coherence,  and Fig.~\ref{fig:phase-flip}(d)  shows the imaginary part of the coherence. The solid lines represent the corresponding theoretical values when the experimental initial state is evolved with the ideal phase flip channel given in Eq.~(\ref{Kraus:phase-flip}).
As can be seen in this figure, the population $H$ and $V$ remain unaltered while the coherence decreases linearly with $q$.  
 
For a more rigourous approach, one should take into account the unavoidable loss of coherence characterized by the parameter $p$, due to imperfections on the SLM. In this case, $p_{\rm{eff}}=q$ is no longer valid. Instead we should use the more general Eq.~(\ref{peff}), replacing $A_{\psi}$ according to Eq.~(\ref{cpf}), for obtaining an effective value $q_{\rm{eff}}$ associated with the degree of decoherence given by:
\begin{eqnarray}
p_{\rm{eff}}=\frac{1-(1-2p)(1-2q)}{2} \equiv q_{\rm{eff}}.
\label{qeff}
\end{eqnarray}
From the above equation we notice that the parameter $p$ sets a lower bound for
the degree of decoherence, now characterized by the parameter $q_{\rm{eff}}$, given in terms of the parameter $q$ that we actually control. This means that the lower value for $q_{\rm{eff}}$ would be $p$.
Assuming that the aimed degree of decoherence characterized by the ideal parameter $q$ in Eq.~(\ref{Kraus:phase-flip}) is greater than this value, this problem is overcome with use of $q_{\rm{eff}}$ given by Eq. (\ref{qeff}).
   
We note that by combining additional wave plates before and after the SLM, the bit flip and the bit phase flip decoherence channels~\cite{NielsenChuang} can also be implemented in a similar fashion.

\section{Conclusion}
 
We use the formalism of quantum channels to characterize the 
action of the SLM on qubits encoded in the polarization degree of freedom of light.
By means of quantum process tomography, we experimentally obtain the Kraus operators that represent the quantum channel describing the effect of the SLM on polarization qubits and propose a theoretical model that matches our experimental results.  As an example of the application of this formalism, we show how a controllable phase flip channel can be implemented, entangling the polarization and the transverse spatial state of light. Considering that little work has been done exploiting this coupling, we expect that our formal characterization scheme will provide incentive for an even richer use of spatial light modulators.

\section{Acknowledgements}

The authors would like to thank Osvaldo Far\'ias, Gabriel Aguilar, and Daniel Tasca for helpful discussions. We acknowledge financial support from the Brazilian funding agencies
CNPq, CAPES, and FAPERJ. This work was performed as part of the Brazilian
Instituto Nacional de Ci{\^e}ncia e Tecnologia -- Informa{\c c}{\~a}o Qu{\^a}ntica (INCT-IQ). G.B.L. is funded by the Austrian Academy of Sciences (\"OAW) through a fellowship of the Vienna Center for Quantum Science and Technology (VCQ).


\begin{thebibliography}{32}
\expandafter\ifx\csname natexlab\endcsname\relax\def\natexlab#1{#1}\fi
\expandafter\ifx\csname bibnamefont\endcsname\relax
  \def\bibnamefont#1{#1}\fi
\expandafter\ifx\csname bibfnamefont\endcsname\relax
  \def\bibfnamefont#1{#1}\fi
\expandafter\ifx\csname citenamefont\endcsname\relax
  \def\citenamefont#1{#1}\fi
\expandafter\ifx\csname url\endcsname\relax
  \def\url#1{\texttt{#1}}\fi
\expandafter\ifx\csname urlprefix\endcsname\relax\def\urlprefix{URL }\fi
\providecommand{\bibinfo}[2]{#2}
\providecommand{\eprint}[2][]{\url{#2}}

\bibitem[{\citenamefont{Moreno et~al.}(2012)\citenamefont{Moreno, Davis,
  Hernandez, Cottrell, and Sand}}]{Moreno:12}
\bibinfo{author}{\bibfnamefont{I.}~\bibnamefont{Moreno}},
  \bibinfo{author}{\bibfnamefont{J.~A.} \bibnamefont{Davis}},
  \bibinfo{author}{\bibfnamefont{T.~M.} \bibnamefont{Hernandez}},
  \bibinfo{author}{\bibfnamefont{D.~M.} \bibnamefont{Cottrell}},
  \bibnamefont{and} \bibinfo{author}{\bibfnamefont{D.}~\bibnamefont{Sand}},
  \bibinfo{journal}{Opt. Express} \textbf{\bibinfo{volume}{20}},
  \bibinfo{pages}{364} (\bibinfo{year}{2012}).

\bibitem[{\citenamefont{Meshulach and Silberberg}(1998)}]{meshulach98}
\bibinfo{author}{\bibfnamefont{D.}~\bibnamefont{Meshulach}} \bibnamefont{and}
  \bibinfo{author}{\bibfnamefont{Y.}~\bibnamefont{Silberberg}},
  \bibinfo{journal}{Nature (London)} \textbf{\bibinfo{volume}{396}},
  \bibinfo{pages}{239} (\bibinfo{year}{1998}).

\bibitem[{\citenamefont{Yao et~al.}(2006)\citenamefont{Yao, Franke-Arnold,
  Courtial, Padgett, and Barnett}}]{Yao:06}
\bibinfo{author}{\bibfnamefont{E.}~\bibnamefont{Yao}},
  \bibinfo{author}{\bibfnamefont{S.}~\bibnamefont{Franke-Arnold}},
  \bibinfo{author}{\bibfnamefont{J.}~\bibnamefont{Courtial}},
  \bibinfo{author}{\bibfnamefont{M.~J.} \bibnamefont{Padgett}},
  \bibnamefont{and} \bibinfo{author}{\bibfnamefont{S.~M.}
  \bibnamefont{Barnett}}, \bibinfo{journal}{Opt. Express}
  \textbf{\bibinfo{volume}{14}}, \bibinfo{pages}{13089} (\bibinfo{year}{2006}).

\bibitem[{\citenamefont{Fatemi et~al.}(2007)\citenamefont{Fatemi, Bashkansky,
  and Dutton}}]{Fatemi:07}
\bibinfo{author}{\bibfnamefont{F.~K.} \bibnamefont{Fatemi}},
  \bibinfo{author}{\bibfnamefont{M.}~\bibnamefont{Bashkansky}},
  \bibnamefont{and} \bibinfo{author}{\bibfnamefont{Z.}~\bibnamefont{Dutton}},
  \bibinfo{journal}{Opt. Express} \textbf{\bibinfo{volume}{15}},
  \bibinfo{pages}{3589} (\bibinfo{year}{2007}).

\bibitem[{\citenamefont{McGloin et~al.}(2003)\citenamefont{McGloin, Spalding,
  Melville, Sibbett, and Dholakia}}]{McGloin:03}
\bibinfo{author}{\bibfnamefont{D.}~\bibnamefont{McGloin}},
  \bibinfo{author}{\bibfnamefont{G.}~\bibnamefont{Spalding}},
  \bibinfo{author}{\bibfnamefont{H.}~\bibnamefont{Melville}},
  \bibinfo{author}{\bibfnamefont{W.}~\bibnamefont{Sibbett}}, \bibnamefont{and}
  \bibinfo{author}{\bibfnamefont{K.}~\bibnamefont{Dholakia}},
  \bibinfo{journal}{Opt. Express} \textbf{\bibinfo{volume}{11}},
  \bibinfo{pages}{158} (\bibinfo{year}{2003}).

\bibitem[{\citenamefont{Curtis et~al.}(2002)\citenamefont{Curtis, Koss, and
  Grier}}]{Curtis2002169}
\bibinfo{author}{\bibfnamefont{J.~E.} \bibnamefont{Curtis}},
  \bibinfo{author}{\bibfnamefont{B.~A.} \bibnamefont{Koss}}, \bibnamefont{and}
  \bibinfo{author}{\bibfnamefont{D.~G.} \bibnamefont{Grier}},
  \bibinfo{journal}{Optics Communications} \textbf{\bibinfo{volume}{207}},
  \bibinfo{pages}{169} (\bibinfo{year}{2002}).

\bibitem[{\citenamefont{Grier}(2003)}]{grier2003}
\bibinfo{author}{\bibfnamefont{D.~G.} \bibnamefont{Grier}},
  \bibinfo{journal}{Nature (London)} \textbf{\bibinfo{volume}{424}},
  \bibinfo{pages}{810} (\bibinfo{year}{2003}).

\bibitem[{\citenamefont{Barreto~Lemos et~al.}(2012)\citenamefont{Barreto~Lemos,
  Gomes, Walborn, Souto~Ribeiro, and Toscano}}]{lemos2012}
\bibinfo{author}{\bibfnamefont{G.}~\bibnamefont{Barreto~Lemos}},
  \bibinfo{author}{\bibfnamefont{R.~M.} \bibnamefont{Gomes}},
  \bibinfo{author}{\bibfnamefont{S.~P.} \bibnamefont{Walborn}},
  \bibinfo{author}{\bibfnamefont{P.~H.} \bibnamefont{Souto~Ribeiro}},
  \bibnamefont{and} \bibinfo{author}{\bibfnamefont{F.}~\bibnamefont{Toscano}},
  \bibinfo{journal}{Nature Communications} \textbf{\bibinfo{volume}{3}},  \bibinfo{pages}{1211}
  (\bibinfo{year}{2012}).

\bibitem[{\citenamefont{Treps et~al.}(2002)\citenamefont{Treps, Andersen,
  Buchler, Lam, Ma\^itre, Bachor, and Fabre}}]{PhysRevLett.88.203601}
\bibinfo{author}{\bibfnamefont{N.}~\bibnamefont{Treps}},
  \bibinfo{author}{\bibfnamefont{U.}~\bibnamefont{Andersen}},
  \bibinfo{author}{\bibfnamefont{B.}~\bibnamefont{Buchler}},
  \bibinfo{author}{\bibfnamefont{P.~K.} \bibnamefont{Lam}},
  \bibinfo{author}{\bibfnamefont{A.}~\bibnamefont{Ma\^itre}},
  \bibinfo{author}{\bibfnamefont{H.-A.} \bibnamefont{Bachor}},
  \bibnamefont{and} \bibinfo{author}{\bibfnamefont{C.}~\bibnamefont{Fabre}},
  \bibinfo{journal}{Phys. Rev. Lett.} \textbf{\bibinfo{volume}{88}},
  \bibinfo{pages}{203601} (\bibinfo{year}{2002}).

\bibitem[{\citenamefont{Lassen et~al.}(2007)\citenamefont{Lassen, Delaubert,
  Janousek, Wagner, Bachor, Lam, Treps, Buchhave, Fabre, and
  Harb}}]{PhysRevLett.98.083602}
\bibinfo{author}{\bibfnamefont{M.}~\bibnamefont{Lassen}},
  \bibinfo{author}{\bibfnamefont{V.}~\bibnamefont{Delaubert}},
  \bibinfo{author}{\bibfnamefont{J.}~\bibnamefont{Janousek}},
  \bibinfo{author}{\bibfnamefont{K.}~\bibnamefont{Wagner}},
  \bibinfo{author}{\bibfnamefont{H.-A.} \bibnamefont{Bachor}},
  \bibinfo{author}{\bibfnamefont{P.~K.} \bibnamefont{Lam}},
  \bibinfo{author}{\bibfnamefont{N.}~\bibnamefont{Treps}},
  \bibinfo{author}{\bibfnamefont{P.}~\bibnamefont{Buchhave}},
  \bibinfo{author}{\bibfnamefont{C.}~\bibnamefont{Fabre}}, \bibnamefont{and}
  \bibinfo{author}{\bibfnamefont{C.~C.} \bibnamefont{Harb}},
  \bibinfo{journal}{Phys. Rev. Lett.} \textbf{\bibinfo{volume}{98}},
  \bibinfo{pages}{083602} (\bibinfo{year}{2007}).

\bibitem[{\citenamefont{Svozil{\'\i}k et~al.}(2012)\citenamefont{Svozil{\'\i}k,
  Le\'on-Montiel, and Torres}}]{PhysRevA.86.052327}
\bibinfo{author}{\bibfnamefont{J.}~\bibnamefont{Svozil{\'\i}k}},
  \bibinfo{author}{\bibfnamefont{R.~d.~J.} \bibnamefont{Le\'on-Montiel}},
  \bibnamefont{and} \bibinfo{author}{\bibfnamefont{J.~P.}
  \bibnamefont{Torres}}, \bibinfo{journal}{Phys. Rev. A}
  \textbf{\bibinfo{volume}{86}}, \bibinfo{pages}{052327}
  (\bibinfo{year}{2012}).

\bibitem[{\citenamefont{Abouraddy et~al.}(2012)\citenamefont{Abouraddy,
  Giuseppe, Yarnall, Teich, and Saleh}}]{Abouraddy-slm-2012}
\bibinfo{author}{\bibfnamefont{A.~F.} \bibnamefont{Abouraddy}},
  \bibinfo{author}{\bibfnamefont{G.~D.} \bibnamefont{Giuseppe}},
  \bibinfo{author}{\bibfnamefont{T.~M.} \bibnamefont{Yarnall}},
  \bibinfo{author}{\bibfnamefont{M.~C.} \bibnamefont{Teich}}, \bibnamefont{and}
  \bibinfo{author}{\bibfnamefont{B.~E.~A.} \bibnamefont{Saleh}},
  \bibinfo{journal}{Phys. Rev. A} \textbf{\bibinfo{volume}{86}},
  \bibinfo{pages}{050303} (\bibinfo{year}{2012}).

\bibitem[{\citenamefont{Fickler et~al.}(2012)\citenamefont{Fickler, Lapkiewicz,
  Plick, Krenn, Schaeff, Ramelow, and Zeilinger}}]{Fickler02112012}
\bibinfo{author}{\bibfnamefont{R.}~\bibnamefont{Fickler}},
  \bibinfo{author}{\bibfnamefont{R.}~\bibnamefont{Lapkiewicz}},
  \bibinfo{author}{\bibfnamefont{W.~N.} \bibnamefont{Plick}},
  \bibinfo{author}{\bibfnamefont{M.}~\bibnamefont{Krenn}},
  \bibinfo{author}{\bibfnamefont{C.}~\bibnamefont{Schaeff}},
  \bibinfo{author}{\bibfnamefont{S.}~\bibnamefont{Ramelow}}, \bibnamefont{and}
  \bibinfo{author}{\bibfnamefont{A.}~\bibnamefont{Zeilinger}},
  \bibinfo{journal}{Science} \textbf{\bibinfo{volume}{338}},
  \bibinfo{pages}{640} (\bibinfo{year}{2012}).

\bibitem[{\citenamefont{Etcheverry et~al.}(2013)\citenamefont{Etcheverry,
  Canas, Gomez, Nogueira, Saavedra, Xavier, and
  Lima}}]{lima-slm-teleportation2013}
\bibinfo{author}{\bibfnamefont{S.}~\bibnamefont{Etcheverry}},
  \bibinfo{author}{\bibfnamefont{G.}~\bibnamefont{Canas}},
  \bibinfo{author}{\bibfnamefont{E.~S.} \bibnamefont{Gomez}},
  \bibinfo{author}{\bibfnamefont{W.~A.~T.} \bibnamefont{Nogueira}},
  \bibinfo{author}{\bibfnamefont{C.}~\bibnamefont{Saavedra}},
  \bibinfo{author}{\bibfnamefont{G.~B.} \bibnamefont{Xavier}},
  \bibnamefont{and} \bibinfo{author}{\bibfnamefont{G.}~\bibnamefont{Lima}},
  \bibinfo{journal}{Scientific Reports} \textbf{\bibinfo{volume}{3}},
  \bibinfo{pages}{2316} (\bibinfo{year}{2013}).

\bibitem[{\citenamefont{Leach et~al.}(2010)\citenamefont{Leach, Jack, Romero,
  Jha, Yao, Franke-Arnold, Ireland, Boyd, Barnett, and
  Padgett}}]{Leach06082010}
\bibinfo{author}{\bibfnamefont{J.}~\bibnamefont{Leach}},
  \bibinfo{author}{\bibfnamefont{B.}~\bibnamefont{Jack}},
  \bibinfo{author}{\bibfnamefont{J.}~\bibnamefont{Romero}},
  \bibinfo{author}{\bibfnamefont{A.~K.} \bibnamefont{Jha}},
  \bibinfo{author}{\bibfnamefont{A.~M.} \bibnamefont{Yao}},
  \bibinfo{author}{\bibfnamefont{S.}~\bibnamefont{Franke-Arnold}},
  \bibinfo{author}{\bibfnamefont{D.~G.} \bibnamefont{Ireland}},
  \bibinfo{author}{\bibfnamefont{R.~W.} \bibnamefont{Boyd}},
  \bibinfo{author}{\bibfnamefont{S.~M.} \bibnamefont{Barnett}},
  \bibnamefont{and} \bibinfo{author}{\bibfnamefont{M.~J.}
  \bibnamefont{Padgett}}, \bibinfo{journal}{Science}
  \textbf{\bibinfo{volume}{329}}, \bibinfo{pages}{662} (\bibinfo{year}{2010}).

\bibitem[{\citenamefont{D'Ambrosio et~al.}(2013)\citenamefont{D'Ambrosio,
  Cardano, Karimi, Nagali, Santamato, Marrucci, and Sciarrino}}]{MUBS}
\bibinfo{author}{\bibfnamefont{V.}~\bibnamefont{D'Ambrosio}},
  \bibinfo{author}{\bibfnamefont{F.}~\bibnamefont{Cardano}},
  \bibinfo{author}{\bibfnamefont{E.}~\bibnamefont{Karimi}},
  \bibinfo{author}{\bibfnamefont{E.}~\bibnamefont{Nagali}},
  \bibinfo{author}{\bibfnamefont{E.}~\bibnamefont{Santamato}},
  \bibinfo{author}{\bibfnamefont{L.}~\bibnamefont{Marrucci}}, \bibnamefont{and}
  \bibinfo{author}{\bibfnamefont{F.}~\bibnamefont{Sciarrino}},
  \bibinfo{journal}{Sci. Rep.} \textbf{\bibinfo{volume}{3}},  \bibinfo{pages}{2726}
  (\bibinfo{year}{2013}).

\bibitem[{\citenamefont{Gibson et~al.}(2004)\citenamefont{Gibson, Courtial,
  Padgett, Vasnetsov, Pas'ko, Barnett, and Franke-Arnold}}]{Gibson:04}
\bibinfo{author}{\bibfnamefont{G.}~\bibnamefont{Gibson}},
  \bibinfo{author}{\bibfnamefont{J.}~\bibnamefont{Courtial}},
  \bibinfo{author}{\bibfnamefont{M.}~\bibnamefont{Padgett}},
  \bibinfo{author}{\bibfnamefont{M.}~\bibnamefont{Vasnetsov}},
  \bibinfo{author}{\bibfnamefont{V.}~\bibnamefont{Pas'ko}},
  \bibinfo{author}{\bibfnamefont{S.}~\bibnamefont{Barnett}}, \bibnamefont{and}
  \bibinfo{author}{\bibfnamefont{S.}~\bibnamefont{Franke-Arnold}},
  \bibinfo{journal}{Opt. Express} \textbf{\bibinfo{volume}{12}},
  \bibinfo{pages}{5448} (\bibinfo{year}{2004}).

\bibitem[{\citenamefont{Gruneisen et~al.}(2008)\citenamefont{Gruneisen, Miller,
  Dymale, and Sweiti}}]{Gruneisen:08}
\bibinfo{author}{\bibfnamefont{M.~T.} \bibnamefont{Gruneisen}},
  \bibinfo{author}{\bibfnamefont{W.~A.} \bibnamefont{Miller}},
  \bibinfo{author}{\bibfnamefont{R.~C.} \bibnamefont{Dymale}},
  \bibnamefont{and} \bibinfo{author}{\bibfnamefont{A.~M.}
  \bibnamefont{Sweiti}}, \bibinfo{journal}{Appl. Opt.}
  \textbf{\bibinfo{volume}{47}}, \bibinfo{pages}{A32} (\bibinfo{year}{2008}).

\bibitem[{\citenamefont{Moreno et~al.}(2003)\citenamefont{Moreno, Vel\'asquez,
  Fern\'andez-Pousa, S\'anchez-L\'opez, and Mateos}}]{moreno2003}
\bibinfo{author}{\bibfnamefont{I.}~\bibnamefont{Moreno}},
  \bibinfo{author}{\bibfnamefont{P.}~\bibnamefont{Vel\'asquez}},
  \bibinfo{author}{\bibfnamefont{C.~R.} \bibnamefont{Fern\'andez-Pousa}},
  \bibinfo{author}{\bibfnamefont{M.~M.} \bibnamefont{S\'anchez-L\'opez}},
  \bibnamefont{and} \bibinfo{author}{\bibfnamefont{F.}~\bibnamefont{Mateos}},
  \bibinfo{journal}{J. of Appl. Phys.} \textbf{\bibinfo{volume}{94}},
  \bibinfo{pages}{3697} (\bibinfo{year}{2003}).

\bibitem[{\citenamefont{Marrucci et~al.}(2006)\citenamefont{Marrucci, Manzo,
  and Paparo}}]{Marruci2006}
\bibinfo{author}{\bibfnamefont{L.}~\bibnamefont{Marrucci}},
  \bibinfo{author}{\bibfnamefont{C.}~\bibnamefont{Manzo}}, \bibnamefont{and}
  \bibinfo{author}{\bibfnamefont{D.}~\bibnamefont{Paparo}},
  \bibinfo{journal}{Phys. Rev. Lett.} \textbf{\bibinfo{volume}{96}},
  \bibinfo{pages}{163905} (\bibinfo{year}{2006}).

\bibitem[{\citenamefont{Nagali et~al.}(2010)\citenamefont{Nagali, Sansoni,
  Marruccci, Santamato, and Sciarrino}}]{Nagali2010}
\bibinfo{author}{\bibfnamefont{E.}~\bibnamefont{Nagali}},
  \bibinfo{author}{\bibfnamefont{L.}~\bibnamefont{Sansoni}},
  \bibinfo{author}{\bibfnamefont{L.}~\bibnamefont{Marruccci}},
  \bibinfo{author}{\bibfnamefont{E.}~\bibnamefont{Santamato}},
  \bibnamefont{and}
  \bibinfo{author}{\bibfnamefont{F.}~\bibnamefont{Sciarrino}},
  \bibinfo{journal}{Phys. Rev. A} \textbf{\bibinfo{volume}{81}},
  \bibinfo{pages}{052317} (\bibinfo{year}{2010}).

\bibitem[{\citenamefont{Lemos et~al.}(2014)\citenamefont{Lemos, Ribeiro, and
  Walborn}}]{Lemos:14}
\bibinfo{author}{\bibfnamefont{G.~B.} \bibnamefont{Lemos}},
  \bibinfo{author}{\bibfnamefont{P.~H.~S.} \bibnamefont{Ribeiro}},
  \bibnamefont{and} \bibinfo{author}{\bibfnamefont{S.~P.}
  \bibnamefont{Walborn}}, \bibinfo{journal}{J. Opt. Soc. Am. A}
  \textbf{\bibinfo{volume}{31}}, \bibinfo{pages}{704} (\bibinfo{year}{2014}).
  
\bibitem[{\citenamefont{Hor-Meyll et~al.}(2014)\citenamefont{Hor-Meyll,
  Almeida, Lemos, Souto~Ribeiro, and Walborn}}]{hormeyll14}
\bibinfo{author}{\bibfnamefont{M.}~\bibnamefont{Hor-Meyll}},
  \bibinfo{author}{\bibfnamefont{J.~O.} \bibnamefont{Almeida}},
  \bibinfo{author}{\bibfnamefont{G.~B.} \bibnamefont{Lemos}},
  \bibinfo{author}{\bibfnamefont{P.~H.} \bibnamefont{Souto~Ribeiro}},
  \bibnamefont{and} \bibinfo{author}{\bibfnamefont{S.}~\bibnamefont{Walborn}},
  \bibinfo{journal}{Phys. Rev. Lett.} \textbf{\bibinfo{volume}{112}},
  \bibinfo{pages}{053602} (\bibinfo{year}{2014}).

\bibitem[{\citenamefont{Nielsen and Chuang}(2000)}]{NielsenChuang}
\bibinfo{author}{\bibfnamefont{M.~A.} \bibnamefont{Nielsen}} \bibnamefont{and}
  \bibinfo{author}{\bibfnamefont{I.}~\bibnamefont{Chuang}},
  \emph{\bibinfo{title}{Quantum Computation and Quantum Information}}
  (\bibinfo{publisher}{Cambridge University Press},
  \bibinfo{address}{Cambridge, UK}, \bibinfo{year}{2000}).

\bibitem[{\citenamefont{Almeida et~al.}(2007)\citenamefont{Almeida, {de Melo},
  Hor-Meyll, Salles, Walborn, Ribeiro, and Davidovich}}]{almeida07}
\bibinfo{author}{\bibfnamefont{M.~P.} \bibnamefont{Almeida}},
  \bibinfo{author}{\bibfnamefont{F.}~\bibnamefont{{de Melo}}},
  \bibinfo{author}{\bibfnamefont{M.}~\bibnamefont{Hor-Meyll}},
  \bibinfo{author}{\bibfnamefont{A.}~\bibnamefont{Salles}},
  \bibinfo{author}{\bibfnamefont{S.~P.} \bibnamefont{Walborn}},
  \bibinfo{author}{\bibfnamefont{P.~H.~S.} \bibnamefont{Ribeiro}},
  \bibnamefont{and}
  \bibinfo{author}{\bibfnamefont{L.}~\bibnamefont{Davidovich}},
  \bibinfo{journal}{Science} \textbf{\bibinfo{volume}{316}},
  \bibinfo{pages}{579} (\bibinfo{year}{2007}).

\bibitem[{\citenamefont{Jim\'enez~Far\'ias
  et~al.}(2009)\citenamefont{Jim\'enez~Far\'ias, Lombard~Latune, Walborn,
  Davidovich, and Souto~Ribeiro}}]{farias09}
\bibinfo{author}{\bibfnamefont{O.}~\bibnamefont{Jim\'enez~Far\'ias}},
  \bibinfo{author}{\bibfnamefont{C.}~\bibnamefont{Lombard~Latune}},
  \bibinfo{author}{\bibfnamefont{S.~P.} \bibnamefont{Walborn}},
  \bibinfo{author}{\bibfnamefont{L.}~\bibnamefont{Davidovich}},
  \bibnamefont{and} \bibinfo{author}{\bibfnamefont{P.~H.}
  \bibnamefont{Souto~Ribeiro}}, \bibinfo{journal}{Science}
  \textbf{\bibinfo{volume}{324}}, \bibinfo{pages}{1414} (\bibinfo{year}{2009}).

\bibitem[{\citenamefont{Far\'ias et~al.}(2012)\citenamefont{Far\'ias, Aguilar,
  Vald\'es-Hern\'andez, Ribeiro, Davidovich, and Walborn}}]{farias12}
\bibinfo{author}{\bibfnamefont{O.~J.} \bibnamefont{Far\'ias}},
  \bibinfo{author}{\bibfnamefont{G.~H.} \bibnamefont{Aguilar}},
  \bibinfo{author}{\bibfnamefont{A.}~\bibnamefont{Vald\'es-Hern\'andez}},
  \bibinfo{author}{\bibfnamefont{P.~H.~S.} \bibnamefont{Ribeiro}},
  \bibinfo{author}{\bibfnamefont{L.}~\bibnamefont{Davidovich}},
  \bibnamefont{and} \bibinfo{author}{\bibfnamefont{S.~P.}
  \bibnamefont{Walborn}}, \bibinfo{journal}{Phys. Rev. Lett.}
  \textbf{\bibinfo{volume}{109}}, \bibinfo{pages}{150403}
  (\bibinfo{year}{2012}).

\bibitem[{\citenamefont{Fickler et~al.}(2013)\citenamefont{Fickler, Lapkiewicz,
  Ramelow, and Zeilinger}}]{Fickler13}
\bibinfo{author}{\bibfnamefont{R.}~\bibnamefont{Fickler}},
  \bibinfo{author}{\bibfnamefont{R.}~\bibnamefont{Lapkiewicz}},
  \bibinfo{author}{\bibfnamefont{S.}~\bibnamefont{Ramelow}}, \bibnamefont{and}
  \bibinfo{author}{\bibfnamefont{A.}~\bibnamefont{Zeilinger}},
  \bibinfo{journal}{arXiv:1312.1306 [quant-ph]}  (\bibinfo{year}{2013}).

\bibitem[{\citenamefont{Alber et~al.}(2002)\citenamefont{Alber, Beth,
  Horodecki, Horodecki, Horodecki, Rötteler, Weinfurter, Werner, and
  Zeilinger}}]{Alberwerner}
\bibinfo{author}{\bibfnamefont{G.}~\bibnamefont{Alber}},
  \bibinfo{author}{\bibfnamefont{T.}~\bibnamefont{Beth}},
  \bibinfo{author}{\bibfnamefont{M.}~\bibnamefont{Horodecki}},
  \bibinfo{author}{\bibfnamefont{P.}~\bibnamefont{Horodecki}},
  \bibinfo{author}{\bibfnamefont{R.}~\bibnamefont{Horodecki}},
  \bibinfo{author}{\bibfnamefont{M.}~\bibnamefont{Rötteler}},
  \bibinfo{author}{\bibfnamefont{H.}~\bibnamefont{Weinfurter}},
  \bibinfo{author}{\bibfnamefont{R.}~\bibnamefont{Werner}}, \bibnamefont{and}
  \bibinfo{author}{\bibfnamefont{A.}~\bibnamefont{Zeilinger}},
  \emph{\bibinfo{title}{Quantum Information: An Introduction to Basic
  Theoretical Concepts and Experiments Series}} (\bibinfo{publisher}{Springer},
  \bibinfo{address}{Berlin}, \bibinfo{year}{2002}).

\bibitem[{\citenamefont{Walborn et~al.}(2010)\citenamefont{Walborn, Monken,
  P\'adua, and Ribeiro}}]{walborn10}
\bibinfo{author}{\bibfnamefont{S.~P.} \bibnamefont{Walborn}},
  \bibinfo{author}{\bibfnamefont{C.~H.} \bibnamefont{Monken}},
  \bibinfo{author}{\bibfnamefont{S.}~\bibnamefont{P\'adua}}, \bibnamefont{and}
  \bibinfo{author}{\bibfnamefont{P.~H.~S.} \bibnamefont{Ribeiro}},
  \bibinfo{journal}{Phys. Rep.} \textbf{\bibinfo{volume}{495}},
  \bibinfo{pages}{87} (\bibinfo{year}{2010}).

\bibitem[{\citenamefont{Wootters}(1998)}]{wooter1998}
\bibinfo{author}{\bibfnamefont{W.~K.} \bibnamefont{Wootters}},
  \bibinfo{journal}{Physical Review Letters} \textbf{\bibinfo{volume}{80}},
  \bibinfo{pages}{2245} (\bibinfo{year}{1998}).

\bibitem[{\citenamefont{Rungta et~al.}(2001)\citenamefont{Rungta,
  Bu\ifmmode~\check{z}\else \v{z}\fi{}ek, Caves, Hillery, and
  Milburn}}]{PhysRevA.64.042315}
\bibinfo{author}{\bibfnamefont{P.}~\bibnamefont{Rungta}},
  \bibinfo{author}{\bibfnamefont{V.}~\bibnamefont{Bu\ifmmode~\check{z}\else
  \v{z}\fi{}ek}}, \bibinfo{author}{\bibfnamefont{C.~M.} \bibnamefont{Caves}},
  \bibinfo{author}{\bibfnamefont{M.}~\bibnamefont{Hillery}}, \bibnamefont{and}
  \bibinfo{author}{\bibfnamefont{G.~J.} \bibnamefont{Milburn}},
  \bibinfo{journal}{Phys. Rev. A} \textbf{\bibinfo{volume}{64}},
  \bibinfo{pages}{042315} (\bibinfo{year}{2001}).

\end{thebibliography}
\end{document}